\overfullrule=0pt
\magnification=1100
\baselineskip=3.1ex
\raggedbottom
\font\eightpoint=cmr8
\font\bcal=cmbsy10
\font\fivepoint=cmr5
\headline={\hfill{\fivepoint EHLMLHS \ 6 Aug, 1996}}
\def\lanbox{\hbox{$\, \vrule height 0.25cm width 0.25cm depth 0.01cm
\,$}}
\def\uprho{{\raise 1pt\hbox{$\rho$}}}
\def\C{{\bf C}}

\def\R{{\bf R}}

\centerline{{\bcal STABILITY} \ {\bcal OF} \ {\bcal RELATIVISTIC}
\ {\bcal MATTER } {\bcal VIA} \ {\bcal THOMAS}{\bf --}{\bcal FERMI} 
\ {\bcal THEORY} }
\bigskip
\bigskip
{\baselineskip=2.5ex
\vfootnote{}{\eightpoint  \copyright 1996 by the authors.
Reproduction of this article, in its entirety, by any means is permitted
for non-commercial purposes.}
\centerline{Elliott H. Lieb\footnote{$^{(1)}$}{\eightpoint Work
partially
supported by U.S. National Science Foundation grant PHY 95-13072.}}
\centerline{Departments of Mathematics and Physics}
\centerline{Princeton University}
\centerline{P.O. Box 708, Princeton, NJ  08544-0708, USA}
\bigskip
\centerline{Michael Loss\footnote{$^{(2)}$}{\eightpoint Work partially
supported by U.S. National Science Foundation grant DMS95-00840.}}
\centerline{School of Mathematics}
\centerline{Georgia Institute of Technology}
\centerline{Atlanta, GA  30332-0160, USA}}
\bigskip
\centerline{Heinz Siedentop\footnote{$^{(3)}$}{\eightpoint Work partially
supported by European Union, grant ERBFMRXCT960001}}
\centerline{Matematisk institutt }
\centerline{Universitetet i Oslo}
\centerline{Postboks 1053}
\centerline{ N-0316 Oslo, Norway}
\bigskip
\bigskip
\bigskip
\noindent
This article is dedicated to our colleagues, teachers, and coauthors
Klaus Hepp and Walter Hunziker on the occasion of their sexagesimal
birthdays. Their enthusiasm for quantum mechanics as an unending source
of interesting physics and mathematics has influenced many.

\bigskip
\bigskip
\bigskip\noindent
{\bf ABSTRACT:} A Thomas-Fermi-Weizs\"acker type theory is
constructed, by means of which we are able to give a relatively simple
proof of the stability of relativistic matter. Our procedure has the
advantage over previous ones  in that  the critical value of the fine
structure constant, $\alpha$,  is raised to 0.77 (recall that the
critical value is known to be less than 2.72). When $\alpha =1/137$,
the largest nuclear charge is 59 (compared to the known optimum value
87). Apart from this, our method is simple, for it parallels the
original Lieb-Thirring proof of stability of nonrelativistic matter,
and it adds another perspective on the subject.

\vfill\eject
\noindent
{\bf I.  INTRODUCTION}
\smallskip
The `stability of nonrelativistic matter' concerns the $N$-body
Hamiltonian 
$$
H=\sum_{i=1}^N {|p_i |} +\alpha V_{\rm c} \ , \eqno(1.1)
$$
where  $V_{\rm c}$ is the Coulomb potential of $K$ fixed nuclei with
nuclear charge $Ze$, with locations  $R_j$ in $\R^3$, and with $N$
electrons.  In units of the electron charge, $e$,
$$   
V_{\rm c}= -V +W +U \ , \eqno(1.2) 
$$
where 
$$ \eqalignno{
V &:= Z\sum_{i=1}^N\sum_{j=1}^K |x_i-R_j|^{-1} \ , &(1.3)\cr
W&:= \sum_{1 \le i<j \le N} |x_i-x_j|^{-1} \ , &(1.4)\cr
U&:= Z^2 \sum_{1 \le i<j \le K}|R_i-R_j|^{-1} \ . &(1.5)\cr}
$$

As usual $p=-i\nabla$ and $|p| = \sqrt{-\Delta}$, and
the $x_j$ are the electron coordinates. The electrons are assumed to
have $q$ spin states each, $q=2$ being the physical value. This means
that the Hilbert space for the $N$-electron functions is the $N$-fold
antisymmetric tensor product of $L^2(\R^3; \C^q)$.
The constant $\alpha=e^2/\hbar c$ is called the fine structure constant.

We can easily include a magnetic field, which means replacing $|p_i|$
by $|p_i + A(x_i)| $. The vector field, $A$, is the vector potential
(in suitable units) of a magnetic field, $A={\rm curl} B$, and can be
arbitrary, as far as the present work is concerned.  A mass can be
included as well, i.e., $|p_i + A(x_i)| $ can be replaced by
$\sqrt{|p_i + A(x_i)|^2 +m^2}-m$.  The inclusion of a mass or magnetic
field, while it changes the energy, does not affect stability. The
reason  for this and the requisite changes will be pointed out in the
final section. It is for simplicity and clarity that we set $m=0$ and
$A=0$.

`Stability of matter' means that the operator, $H$, is bounded 
below by a universal constant times $N+K$, independent of the $R_j$ and
$A$. In our case, because everything scales as an inverse length, 
the lower bound for $H$ is either $-\infty $ or $0$. Thus, we have to 
find the conditions under which $H$ is a positive 
operator. 

Many people worked on various aspects of this problem, including
J.~Conlon (who gave the first proof [C84]), I. Daubechies,
C.~Fefferman, I.~Herbst, T.~Kato, E.~Lieb, R.~de la Llave, R.~Weder, and
H-T.~Yau. A careful, and still current, review of the history is
contained in the introduction to [LY88], to which we refer the reader.
For  present purposes it suffices to note the current state of affairs
concerning the best available constants needed for stability, as
derived in [LY88].  We can list these in a sequence of remarks as
follows:
\smallskip
\item{1.} Stability for any given values, $\alpha_*$ and $Z_*$
implies  stability for all
$0\leq \alpha < \alpha_*$ and $Z< Z_*$. In fact, we can allow the
nuclei to
have different charges $Z_i$, $1\leq i \leq K$, provided $Z_i \leq Z_*$
for all $i$. This follows from some simple concavity considerations and
has nothing to do with the nature of the proof leading to 
$\alpha_*$ and $Z_*$.
\smallskip
\item{2.} Theorem 2 of [LY88] has the strongest results, but it
is limited to the case of zero magnetic field, $A=0$.
The result is that stability occurs if
$$
q\alpha \leq 1/47 \ \ \ {\sl and}\ \ \ Z\alpha \leq 2/\pi \ .\eqno (1.6)
$$
It is not clear to us how to incorporate a magnetic field in the
proof of Theorem 2, and we leave this as an open problem.
\smallskip
\item{3.} Theorem 1 of [LY88] has weaker results, but a simpler proof. 
That proof generalizes easily to  the $A\neq 0$ case, as  pointed out 
in [LLS95]. The result is complicated to state in full generality, 
but a representative example is that stability holds if 
$$
q\alpha \leq 0.032 \ \ \ {\sl and}\ \ \ Z\alpha \leq 1/\pi \ .\eqno (1.7)
$$
It is possible to let $Z\alpha \rightarrow 2/\pi$ at the expense of
$q\alpha \rightarrow 0$.
\smallskip
\item{4.} Instability definitely occurs if $Z\alpha \geq 2/\pi$,
or if $Z_i\alpha \geq 2/\pi$ for any $i$. It 
also occurs if 
$$
\alpha > 128/(15 \pi) \approx 2.72        \eqno (1.8)
$$
for {\it any} positive value of $Z$ and any value of $q$. 
In other words, if $\alpha > 128/(15 \pi)$ and if 
$Z>0$  then one can produce collapse with only one electron, $N=1$, 
by utilizing sufficiently many nuclei, i.e., by choosing $K$
sufficiently large. 
\smallskip
\item{5.} Instability also definitely occurs if ([LY88], Theorem 4)
$$
\alpha > 36 q^{-1/3} Z^{-2/3} \ ,    \eqno (1.9)
$$
which implies that bosonic matter (which can always be thought of
as fermionic matter with $q=N$) is {\it always} unstable. (Note: there
is a typographical error in Theorem 4 of [LY88].)
\bigskip\bigskip

\noindent
{\bf II.  MAIN RESULTS}
\smallskip

The proof of the stability of {\it non}relativistic matter in [LT75]
uses  a series of inequalities to relate the ground state energy of the
Hamiltonian to  the Thomas-Fermi energy of the electron density,
$\uprho(x)$. The chief point is the kinetic energy inequality for an
$N$-electron state $\Psi$, namely 
$$
\langle \Psi \bigl| \ \sum_{i=1}^N
\ |p_i|^2 \  \bigr|\Psi \rangle > {\rm const.} \int \uprho^{5/3} \ . 
$$
The same approach will not work in the relativistic case because  the
corresponding inequality [D83] is, for dimensional reasons, 
$$
\langle
\Psi \bigl| \ \sum_{i=1}^N \ |p_i| \  \bigr|\Psi \rangle > {\rm const.}
\int \uprho^{4/3} \ . 
$$ While $\int \uprho^{5/3}$ can control the Coulomb
attraction $-Z\alpha \int \uprho(x) /|x|$, unfortunately $\int
\uprho^{4/3}$ cannot do so. For this reason no attempt seems to have
been made to imitate the proof in [LT75] of stability in the
relativistic case.

However, the Coulomb singularity {\it can} be controlled by a
Weizs\"acker type term, namely $\left ( \sqrt \uprho \ , |p| \ \sqrt
\uprho \right ) $.  The relativistic kinetic energy can, in turn, be
bounded below by a term of this type plus a term of the $\int
\uprho^{4/3}$ type. This and other essential inequalities will be
explained more fully below. With the `Coulomb tooth' now gone, TF
theory with $\int \uprho^{4/3}$ can deal adequately with the rest of
the Coulomb energy (with the aid of the exchange-correlation energy
inequality [LO81], whose remainder term also has the form $\int
\uprho^{4/3}$).

Before going into details, let us state our main results. First, we
define Thomas-Fermi-Weizs\"acker (TFW) theory as follows: The class of
functions (`densities') to be considered, denoted by ${\cal C}$,
consists of those nonnegative functions $\uprho: \R^3\rightarrow \R^+$
such that $\sqrt \uprho$ and
$\sqrt {|p|\uprho}$ have finite $L^2(\R^3)$ norms, i.e., 
$$
{\cal C} = \left\{ \uprho \ : \ \uprho(x)  \geq 0 \ \ {\rm and} \  \ 
\int_{\R^3} (1+|p|) \ |\widehat{\sqrt\uprho}(p)|^2 dp \ <\infty \right 
\}  \ ,   \eqno (2.1)
$$
where $\widehat {\sqrt\uprho}(p):= (2\pi)^{-3/2}\int_{\R^3} 
\exp[-ip\cdot x]\sqrt\uprho(x)dx$ denotes the Fourier transform of
the function $\sqrt\uprho(x)$. 

Next, we define the functional
$$
T(\uprho) := \int_{\R^3} |p| |\widehat {\sqrt\uprho}(p)|^2 dp 
  \equiv \left ( \sqrt \uprho \ , |p| \ \sqrt \uprho \right )\ .
\eqno(2.2)
$$
The TFW functional, with given positive constants $\beta $ and 
$\gamma$, is  then
$$
{\cal E}(\uprho) := \beta T(\uprho)  +
{3\over 4} \ \gamma \int_{\R^3} \uprho^{4/3}(x) dx - \alpha \int_{\R^3}
 V(x) \uprho(x) dx + \alpha D( \uprho \ ,  \uprho) +\alpha U 
\eqno (2.3)
$$
with 
$$ D( \uprho \ ,  \uprho) := (1/2)\int_{\R^3}\int_{\R^3}
\uprho(x)\uprho(y) |x-y|^{-1} dx dy \ .
$$

The quantity of principal interest is the energy
$$
E^{TFW}:= \inf \left\{ {\cal E}(\uprho) \ : \ \uprho \in {\cal C}
\right\}\ .  \eqno (2.4)
$$
This quantity depends on the parameters $\alpha, \beta$  and
$\gamma$ and on the nuclear coordinates, $R_j$. If, however, we 
try to minimize $E$ over all choices of the nuclear coordinates then
the result is either $0$ or $-\infty$, as can be easily seen from the
fact that all the terms in ${\cal E} $ scale, under dilation, as
an inverse length.
\bigskip
\noindent
{\bf THEOREM 1. (Stability of TFW theory).} {\it The TFW energy,
$E^{TFW}$, in (2.4) is nonnegative if
$$
\beta \geq {\pi \over 2} Z\alpha, \ \ \  {\sl and} \ \ \ 
\gamma \geq 4.8158\ Z^{2/3}\alpha
 \eqno(2.5)
$$
On the other hand, if $\beta < (\pi/2)Z\alpha $ then $E = -\infty$
for every choice of the nuclear coordinates.}

For the next theorem we have to define the density corresponding to 
an $N$-body wave function. If $\Psi$ is an antisymmetric function
of $N$ space-spin coordinates, normalized in the usual way, we define
$$
\uprho_{\Psi}(x) := N\ \sum_{1\leq\sigma_1, \dots, \sigma_N \leq q}
\int_{\R^{3(N-1)}} |\Psi(x,\sigma_1; x_2,\sigma_2; \dots; x_N,\sigma_N)
|^2 dx_2\cdots dx_N \ .  \eqno (2.6)
$$
\bigskip

\noindent
{\bf THEOREM 2. (TFW  theory bounds quantum mechanics).} {\it
Let $\Psi$ be any normalized antisymmetric function, with
$\uprho_{\Psi} $ defined in (2.6). Choose
$$
\beta = {\pi \over 2} Z\alpha \ \ {\rm and} \ \  \gamma=
{4\over 3}\left[1.63 q^{-1/3}\left(1-{\pi \over 2} Z\alpha\right) 
-1.68\ \alpha \right] \ .
\eqno (2.7) 
$$
Assume that $\gamma $ is positive. 
Then, with this definition of the TFW functional (2.3),
$$
\langle\Psi| \  H \ |\Psi\rangle \geq {\cal E}(\uprho_{\Psi})\ . 
\eqno(2.8)
$$ }

A corollary of these two theorems is that {\it our Hamiltonian, $H$, in
(1.1) is stable if}
$$
({\pi \over 2})Z + 2.2159\ q^{1/3}Z^{2/3} + 1.0307\ q^{1/3} \leq 
                               1/\alpha \ .  \eqno(2.9)
$$
(Cf. (1.9)) In particular, {\it with $q=2$ for electrons, relativistic
matter is stable if $\alpha < 0.77$ and if $Z$ is not too large. When
$\alpha =1/137$ the allowed $Z$ is 59,} which compares favorably with
the best possible value $87\approx 137(2/\pi)$.  \smallskip We leave it
as a
challenge to improve our method so as to achieve the value $137(
2/\pi)$ (with a magnetic field present).  As noted above, this value
has been achieved in [LY88], but without a magnetic field.  The most
noteworthy point is the large value of the critical fine structure
constant:  $\alpha_{critical} \geq 0.77$ when $q=2$.  \smallskip The
bound in (2.9) is, in some respects, similar to Theorem 1 in [LY88],
but it is far simpler, clearer and gives the correct $q$-dependence of
$\alpha$ (note that (1.9) gives a similar bound in the other
direction). The chief methodological difference is that
Theorem 6 is used in [LY88], which bounds  the Coulomb potential below
by  a one-body potential. Here, we use  the exchange-correlation
inequality (3.9) instead.  We repeat that the results above also hold
with a magnetic field.

It is to be emphasized that our stability result is really contained in
Theorem 2. Theorem 1 only gives a condition for which ${\cal E}(\uprho)
\geq 0$. A better estimate on the TFW functional will, via Theorem 2,
yield a better stability bound.

\bigskip
\noindent
{\bf III.  SOME ESSENTIAL INEQUALITIES}
\smallskip

There are five known inequalities about Coulomb systems that will 
be needed in our proof of our main theorems. We begin by recalling them.
\smallskip
\noindent
{ \it KINETIC ENERGY LOCALIZATION}, [LY88] pp. 186 and 188.
\smallskip

Denote by $\Gamma_j$ the Voronoi cell in $\R^3$ that contains $R_j$, 
i.e., the set
$$
\Gamma_j := \left\{x \in \R^3\ :\ |x-R_j| \leq |x-R_k| \ {\rm for \
all}\ k \right\} \ ,\eqno (3.1)
$$
and let $D_j$ be half the distance of the j-th nucleus to its nearest
neighbor. These $\Gamma_j$ are disjoint, except for their boundaries
and, being the intersection of half-spaces they are convex sets.  The
ball centered at $R_j$ with radius $D_j$ is denoted by $B_j$.
Obviously, $B_j \subset \Gamma_j$.

For any function $f \in L^2(\R^3)$ there is the inequality
$$
(f, |p| \ f) \geq  \sum_{j=1}^K \int_{B_j} |f(x)|^2
\left \{ {2 \over \pi} |x-R_j|^{-1}-{1 \over D_j}Y\left({|x-R_j| 
\over D_j}\right)
\right \} dx \ . \eqno(3.2)
$$
 The function $Y$ is given by
$$
Y(r)={2 \over \pi(1+r)} + {1+3r^2 \over \pi r(1+r^2)} \ln(1+r)
-{1-r^2 \over \pi r(1+r^2)}\ln(1-r) - {4r \over \pi(1+r^2)} \ln r \ .
\eqno(3.3)
$$
Numerically it is found that [LY88] (2.27) 
$$
4\pi \int_0^1 Y(r)^4 r^2 dr < 7.6245 \ . \eqno (3.4)
$$
\medskip
\noindent
{\it RELATIVISTIC KINETIC ENERGY BOUND FOR FERMIONS}, [D83].
\smallskip
Let $\Psi$ and $\uprho_{\Psi}$ be as in (2.6). Then
$$
\langle \Psi \bigl| \ \sum_{i=1}^N \ |p_i| \  \bigr|\Psi \rangle \geq
1.63 q^ {-1/3}\int_{\R^3} \uprho_{\Psi}^{4/3}(x) dx \ . \eqno (3.5)
$$

A generalization of this, of importance if we wish to include a mass,
is
$$
\langle \Psi \bigl| \ \sum_{i=1}^N \ [\sqrt{p_i^2 +m^2} -m ] \ 
\bigr|\Psi \rangle
\geq   {3\over 8}m ^4 C\int_{\R^3} g\left((\uprho_{\Psi}(x)/C)^{1/3}
m^{-1}\right) dx  \ , 
\eqno (3.6)
$$
with $C=0.163q$ ({\sl sic}) and with
$$
g(t) := t(1+t^2)^{1/2} (1+2t^2) -{8\over3}t^3
- \ln\left[t+(1+t^2)^{1/2}\right] \ . \eqno(3.7)
$$

\medskip
\noindent
{\it GENERAL KINETIC ENERGY BOUND}, [C84], p.454, (and [HO77] for the 
nonrelativistic case). The following
bound follows from a judicious application of Schwarz's inequality. 
$$
\langle \Psi \bigl| \ \sum_{i=1}^N \ |p_i| \  \bigr|\Psi \rangle \geq
\left ( \sqrt {\uprho_{\Psi}} \ , |p| \ \sqrt {\uprho_{\Psi}} \right )
\ .  \eqno(3.8)
$$
This bound holds irrespective of the symmetry type of the wave function.
\medskip\noindent
{\it EXCHANGE AND CORRELATION INEQUALITY}, [LO81].  
If $\Psi $ is a normalized $N$-particle wave function 
there is a lower bound on the 
interparticle Coulomb repulsion in terms of its density:
$$
\langle \Psi \bigl| \sum_{1\leq i<j\leq N} |x_i-x_j|^{-1} \bigr|
\Psi \rangle \geq  D(\uprho_{\Psi}, \uprho_{\Psi}) -
1.68 \int_{\R^3} \uprho_{\Psi}^{4/3}(x) dx \ . \eqno(3.9)
$$
(Once again, the antisymmetry of $\Psi$ plays no role in  this
inequality.)

\medskip
\noindent
{\it ELECTROSTATIC INEQUALITY}, [LY88], p.196. First, we define a 
function,
$\Phi$ on $\R^3$ with the aid of the Voronoi cells mentioned above.
In the cell $\Gamma_j$, $\Phi $ equals the electrostatic potential
generated by all the nuclei {\it except } for the nucleus situated in
$\Gamma_j$ itself, i.e., for $x$ in $\Gamma_j$
$$
\Phi(x):= Z\ \sum_{{i=1\atop i\neq j}}^K |x - R_i|^{-1} \ . \eqno(3.10)
$$

If $\nu$ is any bounded Borel measure on $\R^3$ 
(not necessarily positive) then
$$
{1\over 2}\int_{\R^3}\int_{\R^3} |x-y|^{-1} d\nu(x) d\nu(y)
-\int_{\R^3} \Phi (x) d\nu(x) +U
\geq {1 \over 8}Z^2 \sum_{j=1}^K D_j^{-1} \ . \eqno (3.11)
$$

\bigskip
\noindent
{\bf IV.  PROOFS OF THEOREMS 1 AND 2} 
\smallskip

To prove Theorem 1
we take $\beta =\pi Z\alpha/2$ (if $\beta >\pi Z\alpha/2$ we simply
throw away the excess positive quantity). Using (3.2) with $f$
replaced by $\sqrt \uprho$, we have that 
$$
{\cal E}(\uprho) \geq {\cal E}_1(\uprho) +
                       \alpha {\cal E}_2(\uprho)\ , \eqno (4.1)
$$
where, by adding and subtracting a term 
$\int \Phi\uprho$, with $\Phi(x)$ as in (3.10),
$$
{\cal E}_1(\uprho) := {3\over 4} \ \gamma \int_{\R^3} \uprho^{4/3}(x)
dx -\alpha \int_{\R^3} W(x) \uprho(x) dx  +
	\alpha \int_{\R^3} \Phi(x)\uprho(x) dx      \eqno (4.2)
$$
and 
$$
{\cal E}_2(\uprho) :=
D( \uprho \ ,  \uprho) -
\int_{\R^3} \Phi(x)\uprho(x) dx  +  U \ . \eqno (4.3)
$$

The function $W(x)$ is defined as follows: In the Voronoi cell $\Gamma
_j$ it is given by
$$
W(x) := \Phi(x) +\cases{ Z |x-R_j|^{-1}, &if $|x-R_j| > D_j$ \cr
                          \phantom{x,} & \cr
                         (\pi Z/2)D_j^{-1}Y\left(|x-R_j|/
                    D_j\right), &if $|x-R_j| \leq D_j$\ . \cr}
       \eqno(4.4)
$$
Note that while the terms $\pm\int \Phi\uprho$ that appear in (4.2),
(4.3) are merely 'strategic', the presence of the term  $\Phi(x)$ in
(4.4) is properly part of the potential energy of the electron and is
not arbitrary. Actually, this strategy is the easy part of Fenchel's 
duality theorem (see [R70], p.~327). This duality principle was used
by Firsov[F57] (see also [L81]) in connection with Thomas-Fermi theory; 
the full blown 
duality theory is not needed for our purposes, so we omit it.

We can now seek lower bounds for ${\cal E}_1(\uprho)$ and ${\cal E}_2
(\uprho)$ separately. Using H\"older's inequality, for example, one
easily concludes that the absolute minimum of ${\cal E}_1(\uprho)$ is
$$\eqalignno{
{\cal E}_1(\uprho) &\geq - {\alpha^4\over 4\gamma^3}\int_{\R^3} \left[
                 W(x)-\Phi(x) \right]_+^4dx &\cr
     \phantom{x}   &=- {(\alpha Z)^4\over 4\gamma^3}\sum_{j=1}^K  
                    \left({\pi \over 2}\right)^4 \int_{B_j}
                    D_j^{-4}Y\left(|x-R_j|/
                    D_j\right)^4 dx +\int_
            {\Gamma_j \setminus B_j} |x-R_j|^{-4}
                       dx &(4.5)\cr
     \phantom{x}   &\geq - {(\alpha Z)^4\over 4\gamma^3}
                    \left\{\left({\pi \over 2}\right)^4
                   (4\pi)\int_0^1 Y(r)^4 r^2  dr
                     + 3\pi \right\}\sum_{j=1}^K D_j^{-1}&(4.6)\cr 
      \phantom{x}  &> - {(\alpha Z)^4\over 4\gamma^3} 
         \left\{7.6245 \left({\pi \over 2}\right)^4 +3\pi \right\}
         \sum_{j=1}^K D_j^{-1} \ . &(4.7) \cr}
$$
The last formula uses (3.4).  The second integral in (4.5) is evaluated
in (4.6) as $3\pi/D_j$, and the explanation is the following: If we
integrate $|x-R_j|^{-4 }$ over the exterior of $B_j$ we would obtain
$4\pi/D_j$ as the result. However, we know that the Voronoi cell
$\Gamma_j$ lies on one side of the mid-plane defined by the nearest
neighbor nucleus. This means that the integral over $\Gamma_j\setminus
B_j$ is bounded above by the integral 
$$D_j^{-1}\int_1^{\infty} dz \int
_0^{\infty} (2\pi r dr) [r^2+z^2]^{-2} = 3\pi/D_j \ .
$$

The ${\cal E}_2$ term can be bounded using (3.11) with $d\nu(x) =
\uprho(x) dx$. Thus,
$$
{\cal E}_2(\uprho) \geq {Z^2 \over 8}\sum_{j=1}^K D_j^{-1}\ .\eqno(4.8)
$$
Combining (4.1), (4.7) and (4.8) we have proved Theorem 1. \hfill\lanbox

\smallskip
Theorem 2 is proved by splitting the relativistic kinetic energy $|p|$
into $\beta |p|$ and $(1-\beta)|p|$, with the choice $\beta = \pi Z
\alpha/2$. The inequalities (3.5), (3.8) and (3.9) immediately give us
Theorem 2. \hfill\lanbox

\bigskip
\noindent
{\bf V. INCLUSION OF MASS AND MAGNETIC FIELDS}
\smallskip
{\it INCLUSION OF MASS.} We replace $|p| $ by $\sqrt{p^2+m^2}-m$ and,
in the corresponding TFW theory, we replace the right side of (3.5) by
the right side of (3.6). It is not easy to carry out the rest of the
program in closed form with this more complicated function, however.
Moreover, it unfortunately gives a slightly worse constant than before,
even  when we set $m=0$; instead of $1.63 q^{-1/3}$ in (3.5) we now
have $C^{-1/3} \approx 1.37q^{-1/3}$. The new energy will not be
positive in the stability regime, as we had before. Instead, it will be
a negative constant times $N$. This new value accords with stability
and represents the binding energy of the electron-nuclear system.

Another way to deal with the mass is to observe, simply, that
$\sqrt{p^2+m^2}-m >|p|-m $, the effect of which is to add a term $-Nm$
to the energy estimate. This term satisfies the criterion for
stability, but it has the defect that is huge in real-world terms, for
it equals the rest energy of the electron.  \medskip

{\it INCLUSION OF MAGNETIC FIELD.} Theorem 2, with a magnetic field
included, is a consequence of the following two inequalities which
replace (3.6) and (3.8):
$$
\langle \Psi \bigl| \ \sum_{i=1}^N \ [\sqrt{(p_i+A(x_i))^2 +m^2} -m ] \
\bigr|\Psi \rangle \geq   {3\over 8}m ^4 C\int_{\R^3}
g\left((\uprho_{\Psi}(x)/C)^{1/3} m^{-1}\right) dx  \ , \eqno (5.1)
$$
and
$$ 
\langle\Psi| \  \sum_{i=1}^N|p_i+A(x_i)| \ |\Psi\rangle \geq \left (
\sqrt {\uprho_{\Psi}} \ , |p| \ \sqrt {\uprho_{\Psi}} \right ) \ .
\eqno(5.2)
$$

To define $\sqrt{|p+A|^2+m^2}$, note that if $A \in L^2_{loc}(\R^3 ;
\R^3)$, then $f\mapsto \Vert (p+A)f \Vert_2^2$ is a closed quadratic
form with $C^{\infty}_0(\R^3)$ being a form core [K78],
[S79-1],[LS81].  Thus it defines a selfadoint operator and it is then
possible to define $\sqrt{|p+A|^2+m^2}$ via the spectral calculus.

The {\it diamagnetic inequality for the heat kernel} [S79-2] 
is the pointwise inequality
$$
\left|\ \exp\left[-t(p+A)^2\right]f(x)\ \right| \leq
\exp\left[-tp^2\right]|f|(x) \ .\eqno(5.3) $$
Using the formula
$$
e^{-|a|}={1 \over \sqrt{\pi}} \int_0^{\infty}e^{-{a^2/ 4t}} 
{dt \over \sqrt t} \ , \eqno (5.4)
$$
which holds for any real number $a$ (and hence for any selfadjoint
operator), we obtain the diamagnetic inequality for the 'relativistic
heat kernel'
$$
\left|\ \exp\left[-t\sqrt{(p+A)^2+m^2}\right]f(x)\ \right| \leq
\exp\left[-t\sqrt{p^2+m^2}\right]|f|(x) \ .\eqno(5.5)
$$
By using (5.5), and following the proof of (3.6) in [D83] step by step,
we obtain (5.1).
Likewise, (5.5) and the formula
$$
(f,\ \sqrt{(p+A)^2+m^2}\ f)=\lim_{t \to 0}{1\over t}
\left\{\ (f,f)-(f,\ \exp\left[-t\sqrt{(p+A)^2+m^2}\right]f)\ \right\}
\  , \eqno(5.6) $$
yields
$$
(f ,\ |p+A|\ f) \geq  (|f| ,\ |p|\ |f|)\ .   \eqno (5.7)
$$
Formula (5.2) is now an immediate consequence of (5.7).

\bigskip
\noindent
{\bf REFERENCES}
\smallskip
\item{[C84]} Conlon, J.G., {\it The ground state energy of a classical
gas}, Commun. Math. Phys. {\bf 94}, 439-458 (1984).

\item{[D83]} Daubechies, I., {\it An uncertainty principle for fermions
with generalized kinetic energy},  Commun. Math. Phys. {\bf 90},
511-520 (1983).

\item{[F57]} Firsov, O.B., {\it Calculation of the interaction potential
of atoms for small nuclear separations}, Sov. Phys. JETP {\bf 5},
1192-1196 (1957). 

\item{[HO77]} Hoffmann-Ostenhof, M. and Hoffman-Ostenhof, T., {\it
Schr\"odinger inequalities and asymptotic behavior of the electronic
density of atoms and molecules}, Phys. Rev. A {\bf 16}, 1782-1785
(1977).

\item{[K78]} Kato, T.,  {\it Remarks on Schr\"odinger operators with
vector potentials}, Int. Eq. Operator Theory {\bf 1}, 103-113 (1978).

\item{[LS81]} Leinfelder, H., Simader, C., {\it Schr\"odinger operators
with singular magnetic vector potentials}, Math. Z. {\bf 176}, 1-19
(1981).

\item{[L81]} Lieb, E.H. {\it Thomas-Fermi and related theories of atoms
and molecules}, Rev. Mod. Phys. {\bf 53} 603-641 (1981).
Errata, ibid {\bf 54}, 311 (1982). 

\item{[LLS95]} Lieb, E.H., Loss, M. and Solovej, J.P., 
{\it Stability of Matter in
Magnetic Fields}, Phys.~Rev.~Lett. {\bf 75}, 985-989 (1995).

\item{[LO81]} Lieb, E.H. and Oxford, S., {\it Improved lower bound on
the indirect Coulomb energy}, Int. J. Quant. Chem. {\bf 19}, 427-439
(1981).

\item{[LY88]} Lieb, E.H. and Yau, H-T., {\it The stability and
instability of relativistic matter},  Commun. Math. Phys. {\bf 118},
177-213 (1988).

\item{[LT75]} Lieb, E.H., and Thirring, W.E., {\it Bound for the
kinetic energy of fermions which proves the stability of matter}, Phys.
Rev.  Lett. {\bf 35}, 687-689 (1975). Erratum, {\it ibid}, 1116.

\item{[R70]} Rockafellar, R.T., {\sl Convex Analysis}, Princeton
University Press (1970). 

\item{[S79-1]} Simon, B., {\it Maximal and minimal Schr\"odinger
forms}, J. Opt. Theory {\bf 1}, 37-47 (1979).

\item{[S79-2]} Simon, B., {\it Kato's inequality and the comparison of
semigroups}, J. Funct. Anal. {\bf 32}, 97-101 (1979).

\end